%% file: main.tex
\documentclass[pageno]{jpaper}

\usepackage{mathptmx} 

\usepackage{fancyhdr}
\usepackage[normalem]{ulem}
\usepackage{listings}
\usepackage{float}
\usepackage{tabularx}
\usepackage{xargs}                      

\usepackage{authblk}
\usepackage{caption}

\usepackage{authblk}

\newcolumntype{L}{>{\RaggedRight\arraybackslash}X}
\newcolumntype{C}{>{\centering\arraybackslash}X}

\newcommand*{\thead}[1]{\multicolumn{1}{c}{\bfseries #1}}

\DeclareRobustCommand*\circled[1]{\tikz[baseline=(char.base)]{
            \node[shape=circle,white,draw,fill=black,inner sep=1.4pt] (char) {#1};}}

\newcommand{\ignore}[1]{}

\usepackage[disable]{todonotes} 

\newif\ifsubmit
\submittrue
\ifsubmit
    \newcommand{\usure}[1]{}
    \newcommand{\change}[1]{}
    \newcommand{\info}[1]{}
    \newcommand{\improvement}[1]{}
    \newcommandx{\cheng}[2][1=]{}
    \newcommandx{\abdul}[2][1=]{}
    \newcommandx{\simon}[2][1=]{}
    \newcommandx{\junjun}[2][1=]{}
    \newcommandx{\wenmei}[2][1=]{}
\else
	 
	\newcounter{adtodocounter} 
	\newcounter{cltodocounter} 
	\newcounter{jjtodocounter} 
	\newcounter{wmhtodocounter} 
	\newcounter{sgtodocounter} 
    \newcommandx{\unsure}[2][1=]{\todo[linecolor=red,backgroundcolor=red!25,bordercolor=red,#1]{#2}}
    \newcommandx{\change}[2][1=]{\todo[linecolor=blue,backgroundcolor=blue!25,bordercolor=blue,#1]{#2}}
    \newcommandx{\info}[2][1=]{\todo[linecolor=OliveGreen,backgroundcolor=OliveGreen!25,bordercolor=OliveGreen,#1]{#2}}
    \newcommandx{\improvement}[2][1=]{ \marginpar[\todo[linecolor=Plum,backgroundcolor=Plum!25,bordercolor=Plum,#1]{#2}]{}}
    \newcommandx{\thiswillnotshow}[2][1=]{\todo[disable,#1]{#2}}
    \newcommandx{\cheng}[2][1=]{\stepcounter{cltodocounter} \todo[linecolor=red,backgroundcolor=purple!25,bordercolor=purple,#1]{CL(\thecltodocounter): #2}}
    \newcommandx{\abdul}[2][1=]{\stepcounter{adtodocounter} \todo[linecolor=red,backgroundcolor=red!25,bordercolor=red,#1]{AD(\theadtodocounter): #2}}
    \newcommandx{\simon}[2][1=]{\stepcounter{sgtodocounter} \todo[linecolor=OliveGreen,backgroundcolor=OliveGreen!25,bordercolor=OliveGreen,#1]{SG(\thesgtodocounter): #2}}
    \newcommandx{\jinjun}[2][1=]{\stepcounter{jjtodocounter} \todo[linecolor=yellow,backgroundcolor=yellow!25,bordercolor=yellow,#1]{JJ(\thejjtodocounter): #2}}
    \newcommandx{\wenmei}[2][1=]{\stepcounter{wmhtodocounter} \todo[linecolor=Plum,backgroundcolor=Plum!25,bordercolor=Plum,#1]{WMH(\thewmhtodocounter): #2}}
\fi

\usepackage{blindtext}
\usepackage{amsmath}
\usepackage{algorithmicx}
\usepackage{array}

\definecolor{dkgreen}{rgb}{0,0.6,0}
\definecolor{gray}{rgb}{0.5,0.5,0.5}
\definecolor{mauve}{rgb}{0.58,0,0.82}

\usepackage{fancyhdr}
\usepackage{hyperref}

\newcommand{\nameofproject}{\textit{TrIMS}}
\fancypagestyle{firstpage}{
  \fancyhf{}
\setlength{\headheight}{10pt}
  \pagenumbering{arabic}
}

\begin{document}

\title{TrIMS: Transparent and Isolated Model Sharing for Low Latency Deep Learning Inference in Function as a Service Environments}

\author[1]{\normalsize Abdul Dakkak}
\author[1]{\normalsize Cheng Li}
\author[1]{\normalsize Simon Garcia de Gonzalo}
\author[2]{\normalsize Jinjun Xiong}
\author[3]{\normalsize Wen-mei Hwu}

\affil[ ]{\textit {dakkak@illinois.edu,cli99@illinois.edu,grcdgnz2@illinois.edu,jinjun@us.ibm.com,w-hwu@illinois.edu}}

\affil[1]{%
    \small
  Department of Computer Science \\
  University of Illinois, Urbana-Champaign}
\affil[2]{%
    \small
  IBM Thomas J. Watson Research Center \\
  Yorktown Heights, NY}
\affil[3]{%
    \small
  Department of Electrical and Computer Engineering \\
  University of Illinois, Urbana-Champaign}


\date{}
\maketitle

\thispagestyle{empty}

\begin{abstract}
\input{sec/0-abstract.tex}

\end{abstract}

\input{sec/1-intro.tex}
\input{sec/2-background.tex}
\input{sec/3-method.tex}

\input{sec/4-implementation.tex}
\input{sec/5-evaluation.tex}
\input{sec/6-related.tex}

\input{sec/7-conclusion.tex}
\input{sec/9-ack.tex}

\ifsubmit
\else
\listoftodos
\fi

\bibliographystyle{plain}
\bibliography{ref}

\end{document}

%% file: sec/0-abstract.tex
Deep neural networks (DNNs) have become core computation components within low latency Function as a Service (FaaS) prediction pipelines: including image recognition,  object detection, natural language processing, speech synthesis, and personalized recommendation pipelines.
Cloud computing, as the de-facto backbone of modern computing infrastructure for both enterprise and consumer applications, has to be able to handle user-defined pipelines of diverse DNN inference workloads while maintaining  isolation and latency guarantees, and minimizing resource waste.
The current solution for guaranteeing isolation within FaaS is suboptimal --- suffering from ``cold start'' latency.
A major cause of such inefficiency is the need to move large amount of model data within and across servers.
We propose \nameofproject{} as a novel solution to address these issues. 
Our proposed solution consists of a persistent model store across the GPU, CPU, local storage, and cloud storage hierarchy, an efficient resource management layer that provides isolation, and a succinct set of application APIs and container technologies  for easy and transparent integration with FaaS, Deep Learning~(DL) frameworks, and user code.
We demonstrate our solution by interfacing \nameofproject{} with the Apache MXNet framework and demonstrate up to $24\times$ speedup in latency for image classification models and up to $210\times$ speedup  for large models.
We achieve up to $8\times$ system throughput improvement.

%% file: sec/1-intro.tex

\section{Introduction}\label{sec:introduction}

\begin{figure}[t]
  \centering
  \includegraphics[width=0.5\textwidth]
  {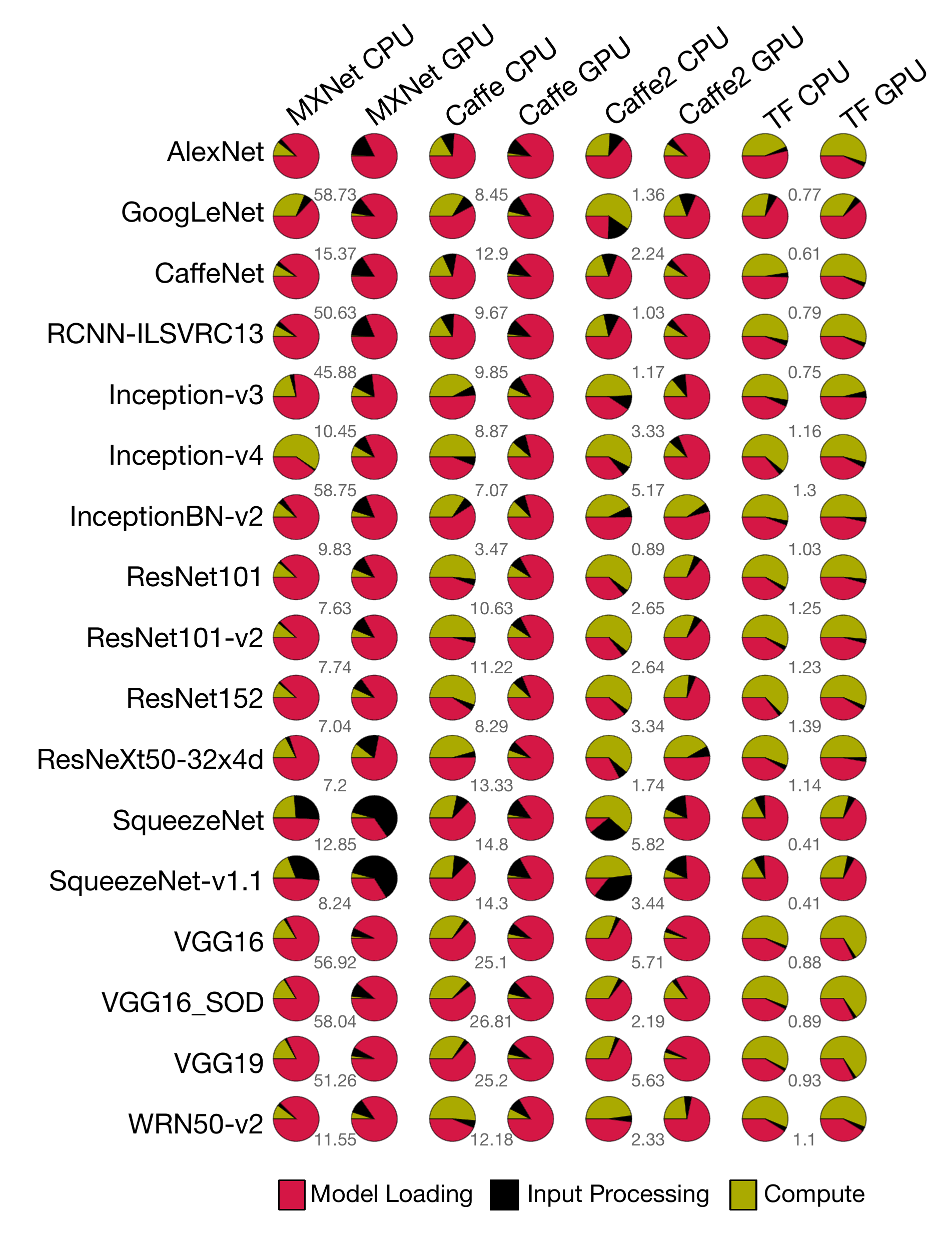}
  \caption{Percentage of time spent in model loading, inference computation, and image preprocessing for ``cold start'' online  DL inference ($batch size = 1$) using CPU and GPU for MXNet, Caffe, Caffe2, and TensorFlow on an IBM S822LC with Pascal GPUs. The speedup of using GPU over CPU for the inference compute alone is shown between the pie charts. Inference time for all frameworks is dominated by model loading except for small models, such as SqueezeNet, where the model size is a few megabytes. For TensorFlow, high GPU initialization overhead impacts the end-to-end time and the achieved speedup.}
  \label{fig:framework_timings_breakdown}
\end{figure}

The recent trend of computing sees a confluence between artificial intelligence, driven primarily by deep learning~(DL), and cloud computing with both gaining traction within enterprise and consumer applications.
Key to this trend is the superior performance, accessibility, and accuracy of deep neural networks~(DNNs) in a wide array of intelligent tasks such as: image recognition, object detection, natural language understanding, speech synthesis, and personalized recommendation.

Today, many business-logic and consumer applications rely on DL inference as core components within their prediction pipelines. 
These pipelines tend to be deployed to the cloud through Function as a Service~(FaaS) platforms ~\cite{googlecloudfuncs, awslambda, azurefuncs, openwhisk}, since they abstract away low-level details such as system setup, dev-ops, and monitoring --- promising service isolation, decentralization, and scalability, while still being more cost-effective compared to dedicated servers.
Since FaaS services execute arbitrary user pipelines, FaaS system \underline{\textbf{must}} execute code in isolation --- through virtual machines (VMs) or containers.

Current off-the-shelf DL inference~\cite{mlaws, awsrek, googlecloudai, azurecog, watson} is performed through HTTP APIs and uses pre-built general models (model catalogs) deployed by the cloud provider or user defined models deployed by the user.
Within the FaaS pipelines, users interact with these models using the HTTP inference APIs and construct their prediction pipelines by defining glue code that parse the input, perform the model prediction, and process the output.
There are two ways to perform model inference, batch prediction and online prediction.
Batch prediction is performed offline on a large set of inputs, while online prediction is usually performed in real-time on a one-by-one basis~\cite{onlinevsbatch, onlinevsbatch2}.
In this paper we focus on online prediction within a latency sensitive FaaS prediction pipeline. 


DL Service providers are aware of the ``cold start'' cost of inference, and therefore eagerly persist models within their catalog --- keeping the models in memory (``warm'') to guarantee the promised latency. 
For example, Amazon ML attempts to respond to most real-time prediction requests within $100ms$~\cite{amazonml}.
Without model persistence, network overhead contributes to a significant portion of the end-to-end inference latency.
As for the user deployed models, the inference latency is not only affected by the network, but is also dominated by the mode inference ``cold start''. 
The ``cold start'' latency can be seconds to minutes depending on the model size and the deployment set-up.
To avoid the ``cold start'' overhead, users have to pay~\cite{googlenodealloc, sagemaker} an hourly cost to persist their models.

FaaS can be used to express latency sensitive prediction pipelines that leverage a chain or ensemble of models.
However, the current practice of integrating FaaS  with model catalogs is inefficient for this usage --- the network latency associated with the inference limits how complex or intelligent a pipeline can be --- making these pipelines out of reach for most but the cloud giants. 
For example, Google Translate targets a $200ms$ per sentence end-to-end latency to avoid user-visible degradation of service~\cite{google-inference}.
To meet the latency requirement, Google implements a monolithic in-house pipeline that uses fast intranet interconnects. Current FaaS users cannot express such a complex pipeline using modular DL inferences to achieve comparable latency.

\begin{figure*}[tp]
  \centering
  \includegraphics[width=\textwidth]{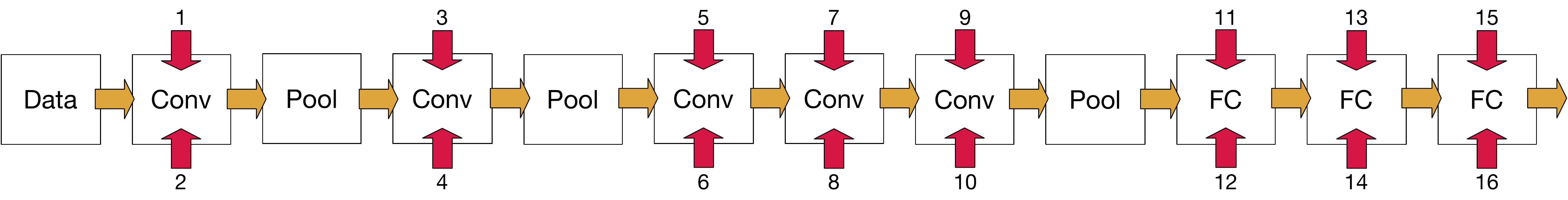}
  \caption{The DL inference graph for AlexNet~\cite{krizhevsky2012imagenet}. The input dimensions and the memory footprint are shown in Table~\ref{tab:alexnet_layers}. }
  \label{fig:inference_architecture}
\end{figure*}

Cloud computing, as the de-facto backbone of modern computing infrastructure, has to be able to enable this scenario in a cost-effective way.
We envision a future FaaS infrastructure that avoids the network overhead, thus making building complex latency sensitive pipelines, with modular DL inference components feasible, while better leveraging the hardware resources.
This enables the development of complex applications based on FaaS; e.g. users can build a personal assistant (similar to Amazon's Alexa or Apple's Siri) by employing off-the-shelf DL inference componets and still achieve comparable latency of the complex monolithic application from cloud giants.
To achieve this goal, we advocate for collocating prediction pipelines with model serving within FaaS, effectively bringing the compute nearer to the model and circumventing the network latency.

The idea of collocating compute with data is not new and has been explored in other domains such as: databases and near memory acceleration.
This paper does not deal with the mechanics of collocating compute and data, since they have been explored elsewhere~\cite{dinsmore2016memory,zheludkov2017high,
lu2013predictive,grandl2016hard,liu2017survey,grandl2017f2}.
Instead, we tackle the challenge faced by collocating model serving and user code within FaaS --- the current method of user code isolation incurs a high ``cold start'' latency for each invocation of the DL inference in the pipeline.



We observe that for ``cold start'' model inference, 
model loading (I/O, data structure deserialization, GPU data movement) is the main source of ``cold start'' latency. 
Figure~\ref{fig:framework_timings_breakdown} shows the ``cold start'' inference time breakdown for popular DL frameworks: Caffe~\cite{jia2014caffe}, Caffe2~\cite{caffe2}, MXNet~\cite{mxnet}, and TensorFlow~\cite{tensorflow-paper-2016}.
For GPU inference, data movement is another contributing factor making GPU less attractive for accelerating inference --- even though GPUs offer a significant compute speed advantage, as shown in Figure~\ref{fig:framework_timings_breakdown}.

We also observe that in a cloud setting DL models are shared extensively across user FaaS pipelines.
For example, Google reported that $41$ natural translation models can accommodate over $75\%$ of their translation requests in~\cite{googlecloudai}. 
Because model parameters are constant, we can leverage model sharing across pipelines by persisting model parameters in GPU and/or CPU memory, hence eliminating the model loading overhead, decreasing the end-to-end latency, and reducing the memory footprint for DL inferences.

In this paper, we propose a \underline{\textbf{Tr}}ansparent and \underline{\textbf{I}}solated \underline{\textbf{M}}odel \underline{\textbf{S}}haring~(\nameofproject{}) scheme to address the ``cold start'' latency challenge faced by collocating user code with model catalogs within FaaS --- it does so while maintaining the isolation constraints, minimizing model-loading overhead, and increasing hardware resource utilization. 
We describe \nameofproject{}'s model resource manager (MRM) which offers a multi-tiered cache for DL models to be shared across user pipelines. 
By decreasing model loading and data movement overhead, \nameofproject{} decreases latency of end-to-end model inference, making inference on GPU a  viable target.
\nameofproject{} also increases memory efficiency for cloud data centers  while maintaining accuracy.


Specifically, this paper makes the following contributions:
\begin{itemize}
    \item We characterize the ``cold start'' overhead for online DL model inference across popular DL frameworks, such as Caffe, Caffe2, MXNet, and TensorFlow, on both CPUs and GPUs and identify model loading as the bottleneck. 
    \item We propose \nameofproject{} to mitigate the model loading overhead faced by collocating user code with model catalogs within FaaS, and increase the model serving efficiency by sharing DL models across all levels of the memory hierarchy in the cloud environment --- GPU, CPU, local storage, and remote storage. To our knowledge, this work is the first to propose sharing DL models across  isolated online prediction pipelines while increasing hardware efficiency and decreasing latency.
    \item We implement \nameofproject{} within Apache MXNet~\cite{mxnet} and evaluate the impact on online inference performance for a representative set of models and systems.
    We show that \nameofproject{} provides $1.12\times$ -- $24\times$ speedup on small (less than 600MB) models and $5\times$ -- $210\times$ speedup on large (up to 6GB)   models and is within $20\%$  of ideal speedup (with ideal being that model loading and data movement taking no time --- i.e. same as persisting the model), and gives $8\times$ system throughput improvement without loss of accuracy.
    \item \nameofproject{} eliminates a substantial part of the non-compute components of the end-to-end latency, making DL model inference on GPU and other novel compute accelerator more viable. We identify remaining latency components for inference, motivating future microarchitecture techniques for further inference latency improvements.


    \item We architect \nameofproject{} so that it can be easily integrated with existing frameworks without user code changes. The method is designed to be compatible with existing framework usage patterns, and requires minimal modifications for framework developers. 
\end{itemize}

The rest of this paper is organized as follows:
Sections~\ref{sec:inference_overhead} and~\ref{sec:faas-pipelines} describes current overheads and practice for inference serving.
Sections~\ref{sec:methodology} and~\ref{sec:implementation} details our design and implementation.
Section~\ref{sec:evaluation} describes our evaluation setup and experiment results.
Section~\ref{sec:related} outlines related work.
Section~\ref{sec:conclusion} concludes.

%



%% file: sec/2-background.tex

\begin{figure*}[t]
  \centering
  \includegraphics[width=\textwidth]{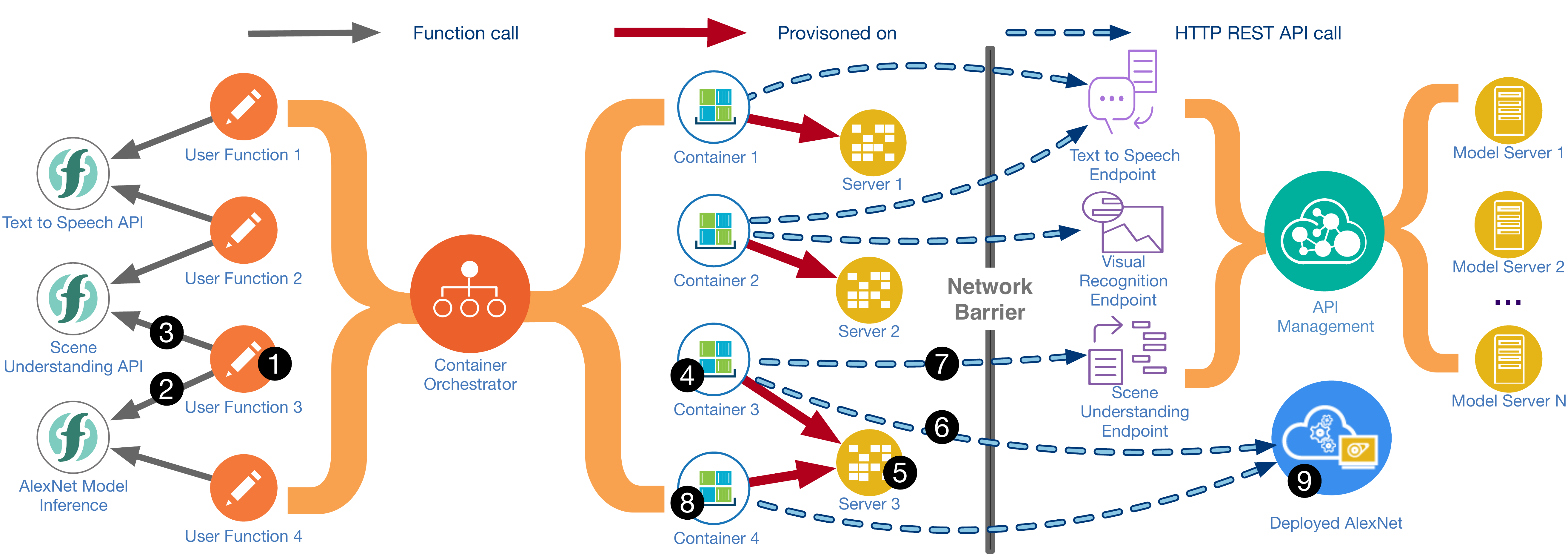}
  \caption{An example of using DL inference in the cloud. \circled{1} application code calls functions from their \circled{2} deployed model or a \circled{3} model catalog. The code is then provisioned onto a \circled{4} container running on \circled{5} server by the cloud provider.
  The code performed API calls to \circled{6} perform AlexNet inference and \circled{7} the scene understanding API. \circled{9} AlexNet is deployed by users through the cloud provider's cloud deployment mechanism.
  } 
  \label{fig:faas_architecture}
\end{figure*}

\section{Deep Learning Inference Overhead}\label{sec:inference_overhead}





\begin{table}[tp!]
    \centering
\begin{tabular}{ | c | l | c | r | }
    \hline
    Index & Name & Dim & MF (MB) \\ 
 \hline
1 & conv1\_bias & $96$  & $0.001$ \\
2 & conv1\_weight & $96 \times 3 \times 11 \times 11$ & $0.270$ \\
3 & conv2\_weight & $256 \times 48 \times 5 \times 5$ & $2.458$ \\
4 & conv2\_bias & $256$ & $0.002$ \\
5 & conv3\_weight & $384 \times 256 \times 3 \times 3$  & $7.078$ \\
6 & conv3\_bias & $384$ & $0.003$ \\
7 & conv4\_bias & $384$ & $0.003$ \\
8 & conv4\_weight & $384 \times 192 \times 3 \times 3$  & $5.3086$ \\
9 & conv5\_weight & $256 \times 192 \times 3 \times 3$  & $3.539$ \\
10  & conv5\_bias & $256$ & $0.002$ \\
11  & fc6\_bias & $4096$  & $0.033$ \\
12  & fc6\_weight & $4096 \times 9216$  & $301.990$ \\
13  & fc7\_weight & $4096 \times 4096$  & $134.218$ \\
14  & fc7\_bias & $4096$  & $0.033$ \\
15  & fc8\_bias & $1000$  & $0.008$ \\
16  & fc8\_weight & $1000 \times 4096$  & $32.768$ \\
     \hline
\end{tabular}
    \caption{Memory footprint (MF) for each layer in Figure~\ref{fig:inference_architecture}.}
    \label{tab:alexnet_layers}
\end{table}


A single DL inference  is much less computationally intensive than training, making it more sensitive to the data loading and deserialization overhead.
A DL inference compute graph is a DAG composed of a set of network layers.
Each computational layer is parameterized through weights and constants.
The model parameters along with the compute topology identify the model~\footnote{Throughout this paper, sharing a layer means that we are sharing both the weights and constants that parameterize the layer.}.
Each layer operator is a function of the incoming edges in the graph and the weights/constants.
An inference pass iterates through the layers of a compute graph and applies the layer operators to its input. Figure~\ref{fig:inference_architecture} shows the inference compute graph for AlexNet~\cite{krizhevsky2012imagenet} and Table~\ref{tab:alexnet_layers} lists the dimension and memory footprint for each layer.

For GPUs, the compute graph and associated weights are  loaded and copied to GPU memory ahead of the computation.
Memory for intermediate layer outputs also need to be allocated.
AlexNet, for example, requires $516MB$ of extra GPU memory to store the intermediate results during the inference process. 
These intermediate outputs are not constant and cannot be shared, since they depend on the user's input.
However, layer weights are constant and can be shared across processes.
For AlexNet, this results in sharing $238MB$ of constant data.

When compute is optimized, the overhead of model loading is magnified.
Figure~\ref{fig:framework_timings_breakdown} shows that GPU outperforms the CPU in terms of compute, thus making model loading a bottleneck for end-to-end inference.
Without data transfer overhead the NVIDIA Tesla V100 GPU using Tensor Cores can achieve $70\times$ higher throughput on CNNs and $130\times$ higher throughput on RNNs compared to a high-end CPU server~\cite{nvidiainfer}.
Reducing the data movement overhead makes GPU a more appealing option for DL inference.

To mitigate the model loading overhead, cloud services and previous work~\cite{olston2017tensorflow,baylor2017tfx, crankshaw2017clipper} persist model catalogs in memory or perform inference in batches.
These strategies require knowledge of the model requests, have potential  resource waste since the system resources are persisted within processes for models even when they are not used, or increase the latency of requests if batching the inferences.

\section{Current Prediction Pipelines in FaaS}\label{sec:faas-pipelines}

Function as a Service (FaaS) is a cost-effective way for users to deploy functions or pipelines that are executed within the cloud.
Users define prediction pipelines that use models they deployed or ones found within the model catalog.
The pipelines are then mapped to a fabric of containers --- used to maintain software stack separation, virtualize system resources, and provide isolation --- that run on physical machines. 
Unlike traditional cloud execution, the functions executed in a FaaS are short lived and are priced on a per-invocation basis (with function execution time and resource utilization being the main cost factors).
Because cloud providers  use a per-API call and per-resource utilization price model, resource waste affects the cloud user's total cost of ownership.

To motivate our work, we use image to scene description pipeline deployed within FaaS as an example --- illustrated in Figure~\ref{fig:faas_architecture}. 
The pipeline takes an image input and outputs a textual description, leveraging a deployed AlexNet and  an off-the-shelf scene understanding model from the cloud provider's model catalog.
Both the AlexNet model inference \circled{2} and the scene understanding API \circled{3} are called within User Function 3 \circled{1}.
Cloud providers then provision the function to run within a container~\circled{4} on a cloud server~\circled{5}.
When user code is triggered, both \circled{6} the deployed AlexNet model and \circled{7} the scene understanding endpoints are  called through HTTP REST API calls.
Meeting the latency requirements for this application is challenging because of the multiple over-the-network requests.

\todo{Because of hardware  resource and cost constraints all the models cannot be loaded on the GPU}

To avoid the network latency, a common practice is to collocate the model within the deployed functions or the application pipelines.
However, such embedding requires a  copy of the model to be loaded privately for each function or application pipeline. 
For example, \circled{4} and \circled{8} have to load \circled{2} AlexNet twice on the same machine --- wasting memory resources. 
The private loads introduces latency overhead, since the model needs to be loaded for the first function invocation.
Since in FaaS isolation must be guaranteed, the previously mentioned persistence schemes, in Section~\ref{sec:inference_overhead}, is not a solution.
Similarly, batching does not apply for low latency inference.


\begin{figure}[h]
  \centering
  \includegraphics[width=0.45\textwidth]{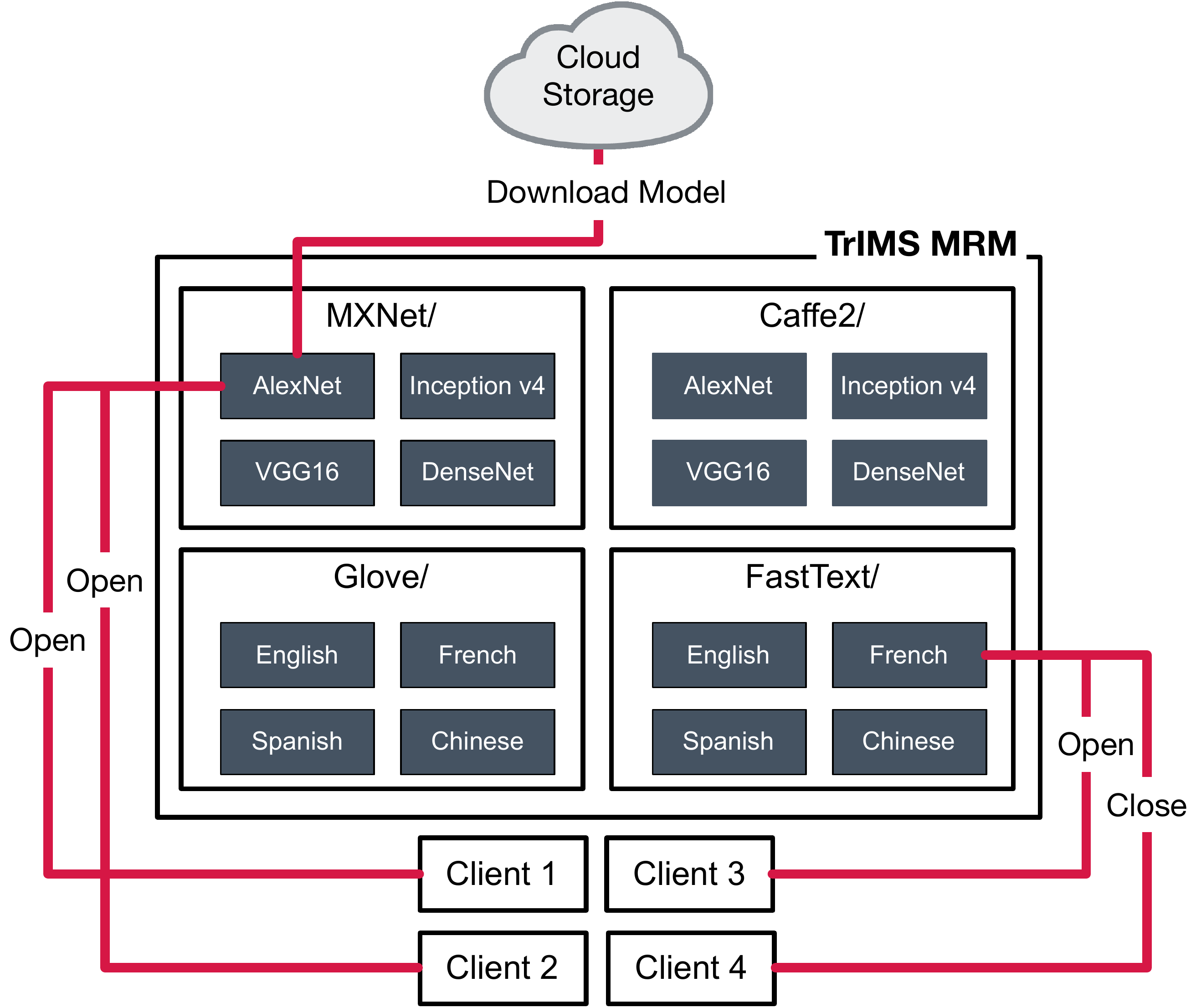}
  \caption{Multiple processes can perform IPC requests to the \nameofproject{} Model Resource Manager (MRM) server; for example $Client_1$, $Client_2$, and $Client_3$ are performing an \texttt{Open} request, while $Client_4$ is performing a \texttt{Close} request. \nameofproject{}'s MRM is responsible for loading and managing the placement of the models in GPU memory, CPU memory, or local disk.}
  \label{fig:trims_architecture}
\end{figure}

 In a cloud setting DL models are shared extensively across user functions, for example: between the $4$  user functions shown in Figure~\ref{fig:inference_architecture}.
 Based on this observation, we propose 
\nameofproject{} to eliminate such model loading overhead and hardware resource waste, while maintaining resource utilization efficiency and decreasing inference latency in user processes.
\nameofproject{} achives this by folding ``private copies'' of the model into a shared copy under the hood.
This is performed by decoupling the model persistence from the user-code execution --- enabling model sharing, isolation, and low latency inference.




%% file: sec/3-method.tex

\section{TrIMS Design}\label{sec:methodology}

\nameofproject{} consists of two components: a Model Resource Manager (MRM) server and framework clients.
MRM manages the model resources resident in the system memory and abstracts away the model loading from framework clients. Each framework client communicates with MRM through inter-process communication~(IPC), as shown in Figure~\ref{fig:trims_architecture}. 
Since  \nameofproject{}  follows the original DL framework's API and semantics --- returning the same  data structures as the unmodified framework --- 
user code can leverage \nameofproject{} transparently without any code modification.

\begin{figure}[b!]
  \centering
  \includegraphics[width=0.45\textwidth]{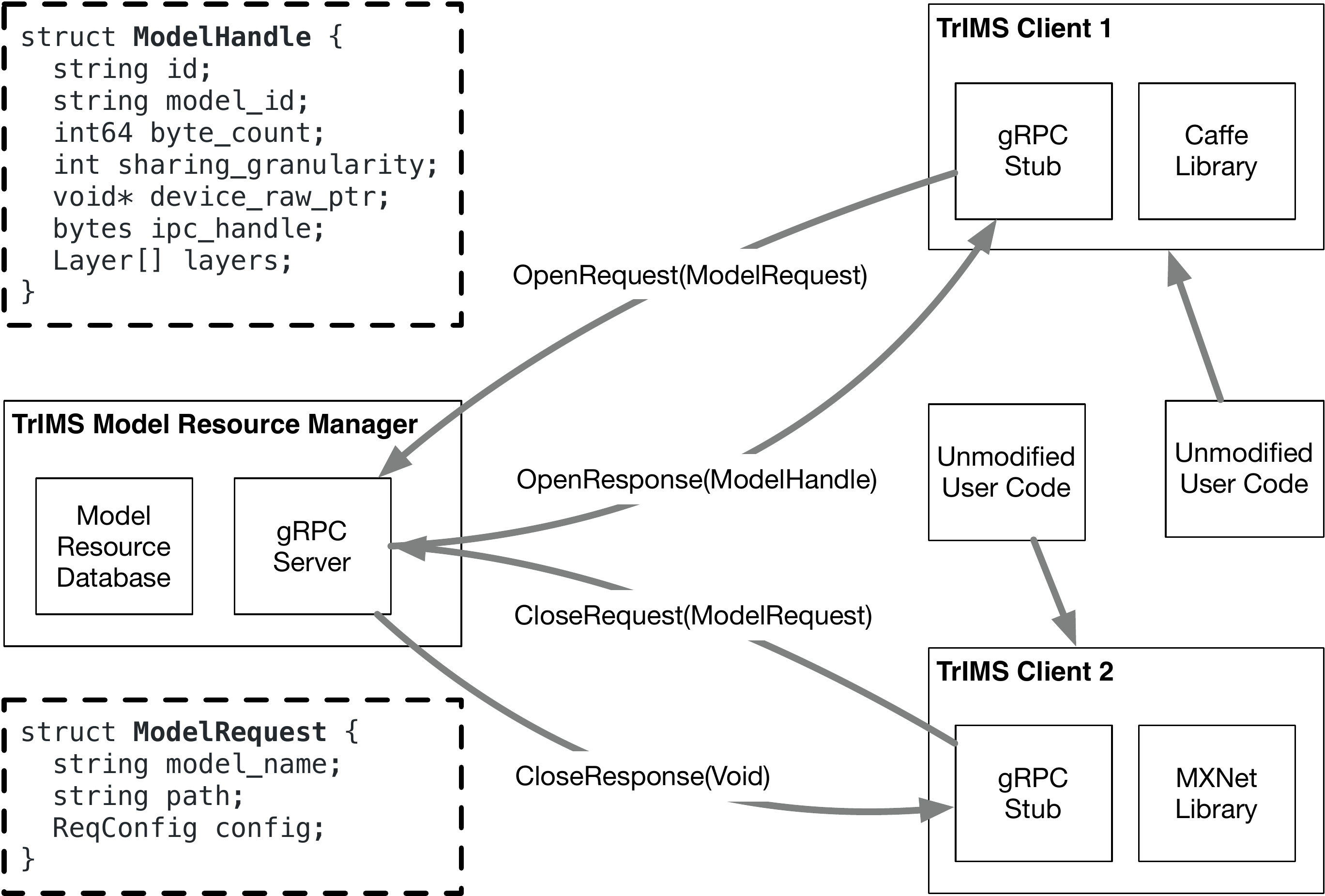}
  \caption{When user code loads a model using the original framework API, instead of loading the model directly from disk, the corresponding \nameofproject{} client sends an \texttt{Open} request with \texttt{ModelRequest} structure to the MRM, and receives a response of type \texttt{ModelHandle}, from which it constructs the compute graph with model weights. When user code unloads a model, then instead of directly destroying the allocated memory, the \nameofproject{} client sends out a \texttt{Close} request with \texttt{ModelHandle} and \nameofproject{} MRM does the housekeeping.}
  \label{fig:trims_api}
\end{figure}

\subsection{\nameofproject{} Model Resource Manager (MRM)}\label{sec:server}

\nameofproject{}'s MRM is a model server daemon that performs model management and placement.
MRM maintains a database of models, addressing them using namespaces, with framework as well as model name and version being used to distinguish frameworks and models.
Figure~\ref{fig:trims_architecture} shows that MRM is managing models for MXNet, Caffe2 DL frameworks as well as word vector embedding models for FastText and Glove.

The MRM placement manager then maps the models into either GPU memory, CPU memory, local storage, or cloud storage.
The four levels are analogous to the traditional CPU cache hierarchy. Because of this,
we will simply refer to these four different memory
hierarchies as ``cache''
in the rest of this paper
whenever there is no ambiguity.

\begin{figure}[b]
  \centering
  \includegraphics[width=0.5\textwidth]{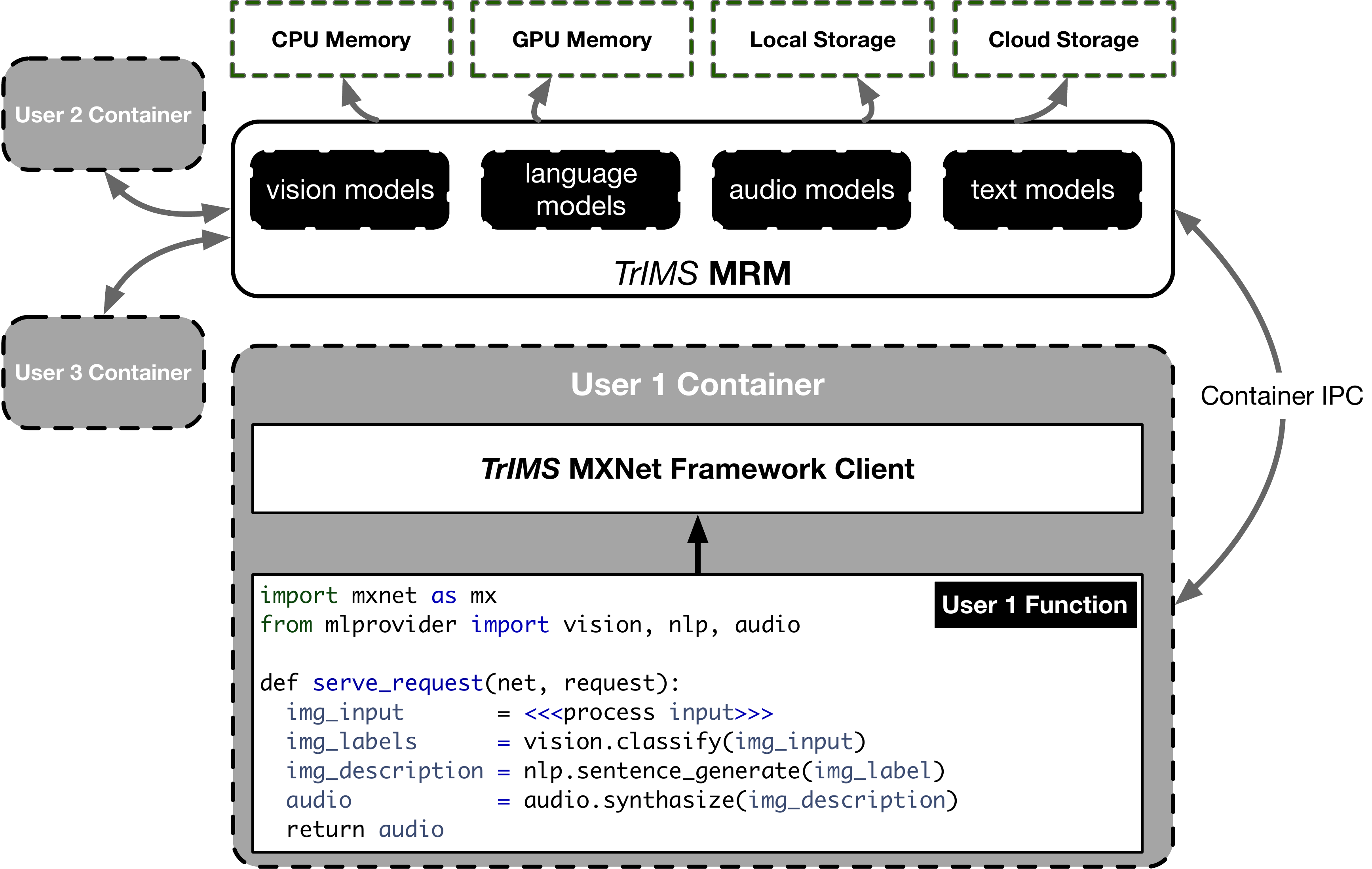}
  \caption{Cloud providers can use \nameofproject{} MRM as a container plugin to provision running untrusted user functions while still leveraging model sharing.  User code is executed within an isolated containers and can get the benefits of \nameofproject{} without code modifications. Sharing occurs when the users utilize the same models as their peers --- which is not uncommon in cloud settings using cloud provided APIs.}
  \label{fig:user_code}
\end{figure}

After system cold boot, initial model requests miss the GPU, CPU, and local storage caches, causing the model to be downloaded from the cloud storage and loaded into the ``caches'' to serve both the current quest and future requests.
When one of the caches becomes full, one or more models are evicted from the cache.

For inter-process communication, \nameofproject{}  uses gRPC~\cite{grpc}  to send and receive messages between the MRM and its clients. 
\nameofproject{} leverages the CUDA runtime's \textit{cudaIpc*} to share GPU memory across processes. 
MRM abstracts away the model management, exposing two API functions to be used by the clients: \texttt{trims::open} and \texttt{trims::close} to load and close a model, respectively.
MRM maintains a reference count for each model to determine the number of users currently using the shared model.
The API is shown in Figure ~\ref{fig:trims_api}.

\begin{figure}[ht!]
    \centering
    \includegraphics[ width=0.5\textwidth]{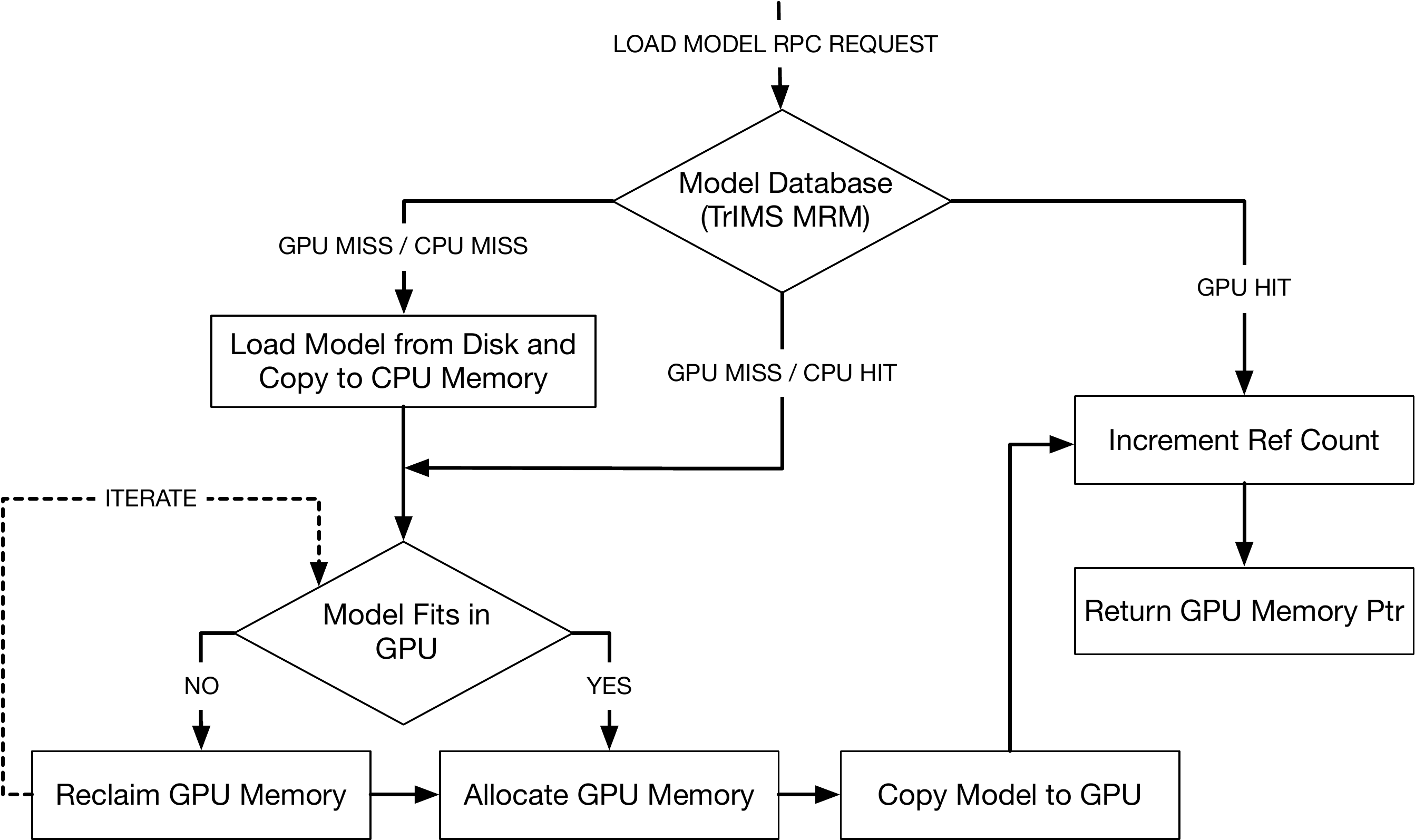}
    \caption{The logic for caching models on both GPU and CPU. The \nameofproject{} client initiates the load model call to \nameofproject{} MRM and gets back a pointer to GPU memory.}
    \label{fig:load_model_flow}
\end{figure}

\subsubsection{Loading Models}

When loading a model, MRM performs shape inference on the model to estimate its memory footprint when running on GPU.
Shape inference is a simple arithmetic computation performed by any framework to determine the amount of internal memory to allocate for a model.
After shape inference, MRM follows the state diagram shown in Figure~\ref{fig:load_model_flow} and needs to handle three cases:

\paragraph{GPU cache hit --- Model is persistent in GPU memory}  
MRM increments the model's reference count and creates a shared memory handle from the device memory owned by MRM. The handle is then returned to the framework client.
Model eviction is triggered when the intermediate results for a model is greater than the available free memory.

\paragraph{GPU cache miss / CPU cache hit --- model is persistent in CPU memory} 
The server queries the current memory utilization of the GPU to see if the model can be copied to GPU memory. If it can, then GPU memory is allocated and copied; if not, then some memory needs to be reclaimed --- entering the memory reclamation procedure.

\paragraph{CPU and GPU cache miss --- model is not persistent in memory}
If the data is not on local storage, then MRM downloads the model from the cloud. If the data is on disk, then MRM loads the data from disk using the framework's serializer. 
Pinned memory is allocated on the CPU and the model weights is copied to it.
MRM then follows the same logic as when the data is persistent in CPU memory.


\subsubsection{Reclaiming Memory and Evicting Models}

Memory reclamation is performed when the memory space for MRM at a specific cache level is full.
Which model to evict to reclaim memory is determined by the eviction policy.
\nameofproject{} supports a pluggable set of common eviction policies such as least recently used(LRU) and least commonly used (LCU).
For the CPU and GPU level caches, one needs to make sure that eviction does not interfere with user's code. 
Models within the MRM database are not candidates for reclamation if they are in use; i.e. the reference count of a model is non-zero.
Evicting models that is currently being used (effectively freeing GPU memory that's being used) causes undefined behavior in the user's code.

\subsubsection{Unloading Models}

When a \nameofproject{} framework client unloads a model (or the user process exists), a model unload request is sent to MRM. 
MRM looks up the model in the database and decrements its reference count.
By default MRM does not free resources for models that have a zero reference count (not currently used), but MRM can be configured to eagerly reclaim these models.

\subsection{\nameofproject{} Frameworks}\label{sec:trims_users}

MRM can handle requests from multiple \nameofproject{}-enabled frameworks, managing their weights (which have different data layouts) in separate namespaces. 
Shown in Figure~\ref{fig:trims_api}, when a \nameofproject{} framework performs a model load request, the framework's name and version are sent along with the request.
The server can then perform the model unmarshaling from disk using the format supported by the framework.

To enable \nameofproject{} in a framework, the functions to load and unload models need to be modified to perform $gRPC$ requests to MRM.
Since, each framework may have its own serialization format, support for the model format, to enable unmarshaling the data from disk to memory, needs to be added to MRM.
With these changes, any type of network supported by the framework (CNN, RNN, Word2Vec, etc.) and any compute pattern is automatically supported by \nameofproject{}.

\subsubsection*{User application rewriting overhead}
---
Since MRM does not modify the framework's API, code that is linked with a \nameofproject{}-enabled framework does not require any change.
\nameofproject{} works within Python, Java, or R.
This is an attractive feature, since the benefits of \nameofproject{} can be leveraged by cloud provider transparently from the user.

\subsubsection*{Sharing Granularity}
---
\nameofproject{} supports fixed-size block, layer, and model level sharing granularity.
Sub-model level sharing granularity is interesting when considering layers or memory across models.
For example, models trained using transfer learning~\cite{torrey2009transfer} share the frozen layer weights.
Block level granularity can also be used to share fixed-size buffers.

\subsubsection*{Multi-GPU and Multi-Node Support}
---
Multi-GPU is usually used when performing batched inference~\cite{tensorflowserving,baylor2017tfx}.
\nameofproject{} inherently supports the multi-GPUs  by leveraging Unified Memory~(UM)~\cite{nvidiaum}.
Support for Multi-GPU sharing can also be performed without relying on UM by making the \nameofproject{} framework client query the device ID of the current GPU context when a model is loaded.
The framework client can then send the device ID along with the request.
\nameofproject{} MRM would then load the model into the GPU with that device ID.
When a request loads a model on a GPU and the requested model is persistent on another GPU, MRM will perform GPU peer-to-peer memory copy if supported.

Multiple independent instances of \nameofproject{} MRM can be loaded for multi-node support and an off-the-shelf task scheduling and load balancing middleware can be used to route and load balance inference requests.
\nameofproject{} can be setup to advertise the models that have already been loaded by users and the current system load to the load balancer.

\subsection{Inference Isolation and Fairness}\label{sec:isolation}


To enable seamless container isolation, \nameofproject{} provides a Docker~\cite{merkel2014docker} volume plugin that allows service providers to provision the container with a communication link to the \nameofproject{} MRM.
The \nameofproject{} MRM process runs in the host system with a link for  frameworks to communicate with it across container boundaries.
Figure~\ref{fig:user_code} shows how untrusted user code can be run on a multi-tenant system while maintaining isolation.
The code shows how users can use DL models, provided by the cloud provider, to create an image to audio pipeline.
The user uses the cloud provided vision, text, and audio models via a library that is part of a model catalog.
All user code executes within a container that communicates with the MRM via the container's IPC mechanism.

%% file: sec/4-implementation.tex

\section{Implementation}\label{sec:implementation}

The experiments reported in this paper are based on an implementation of \nameofproject{} on top of the Apache MXNet~\footnote{The source code for \nameofproject{} is open source and is found at http://github.com/REMOVED/DURING/REVIEW} --- a popular machine learning framework. The \nameofproject{} MRM includes serialization code from MXNet to unmarshal MXNet models from disk. 
We also modify the MXNet framework to integrate it with \nameofproject{} --- keeping the MXNet APIs unchanged. 
Communication between the MXNet framework client and the MRM uses Google's gRPC~\cite{grpc} with the packets encoded using Protocol Buffers~\cite{protobuf}.

To validate the efficiency and generality of our proposal, we follow a few principles throughout our implementation --- even if disregarding some would have given us better speedup:

\subsubsection*{Backward Compatible}
---
The implementation needs to work with the existing framework's code base and language bindings, i.e. we should be able to run preexisting MXNet codes written in Python or Scala with no modifications.
\subsubsection*{Simple and Minimal}
--- The implementation needs to be simple and not modify the framework code as much as possible. Our modifications adds only $1500$ lines of code (less than $0.5\%$ of the MXNet code base) to the framework ($800$ lines for the server and $700$ lines for the client) and is self contained.
\subsubsection*{Configurable} 
--- The implementation has knobs to tweak everything from the eviction strategy of memory sharing, the amount of memory that can be used, whether to enable \nameofproject{}, the levels of cache to enable, etc...
\subsubsection*{Fast, Concurrent and Scalable} 
--- We communicate using gRPC and use efficient data structures~\cite{herlihy2008hopscotch} for the MRM database to make the serving fast and concurrent. The memory sharing strategy in \nameofproject{} is scalable and can handle large amount of load.






\subsection{\nameofproject{} Apache MXNet Framework}\label{sec:framework_client}

We implement \nameofproject{} on top of the Apache MXNet framework client by modifying the \texttt{MXPredCreate} and \texttt{MXPredFree} in the MXNet C predict API's implementations.
When \nameofproject{} is enabled, \texttt{trims::open} and \texttt{trims::close} are called as part of the predictor creation and deletion.
Listing \ref{lst:predictapi} shows the main modification to the original MXNet code.

\begin{lstlisting}[
  float=htp,
  floatplacement=tbp,
  basicstyle=\fontsize{7}{7}\ttfamily,
caption={To integrate \nameofproject{} with MXNet we modify both the \texttt{MXPredCreate} and \texttt{MXPredFree} functions. \texttt{MXPredCreate} loads the model and initializes the compute graph to perform inference, if \nameofproject{} is enabled, we  call \texttt{trims::open} instead of \texttt{NDArray::Load} which loads the model from disk. To correctly free the models, we modify the \texttt{MXPredFree} function to call \texttt{trims::close}. \texttt{MXPredFree} is called in the Predictor destructor or at process exit. },frame=single,label=lst:predictapi,captionpos=b]
MXAPIPredictor MXPredCreate(MXPredParams * p){
  MXAPIPredictor *ret = new MXAPIPredictor();
  {...} // load in the symbol and model parameters
  {... shapes = infer_model_shapes(p); ... }
  if (trims::ENABLED) {
    auto tinfo      = trims::open(...);
    ret->handle_id  = std::get<0>(tinfo);
    ret->model_id   = std::get<1>(tinfo);
    goto setup_predictor;
  }
  // original model loading
  dmlc::MemoryStream fi(p->buf, p->size);
  NDArray::Load(&fi, &data, &names)
  {...}
setup_predictor:  
  {...}
  return ret;
}
void MXPredFree(PredictorHandle handle) {
  auto pred = (MXAPIPredictor *) handle;
  if (trims::ENABLED) trims::close(pred);
  delete pred;
}
\end{lstlisting}

Like most open-source DL Frameworks, MXNet is optimized for training and not inference. We apply a set of optimizations to the original MXNet to improve the inference latency.
The optimizations avoid eager initialization of CUDA resources, remove cuDNN algorithm selection for backward propagation, and simplify the random resource generation. With our optimizations, MXNet is $6\times$ faster for inference on average than the vanilla MXNet for the suite of models we use.
We use the modified MXNet as our baseline for evaluation.

\subsection{GPU Memory Sharing}\label{sec:gpu_share}

We perform GPU memory sharing using the CUDA's \texttt{cudaIPC*} runtime functions.
For Pre-Volta GPUs, the CUDA IPC mechanism utilizes CUDA MPS --- an intermediate user process where the memory allocations are  performed.
This means that all CUDA
operations end up serialized and executed within the same CUDA MPS context --- enabling difference processes to share the same GPU virtual address space (VAS).
For Volta GPUs, NVIDIA introduced a new feature to allows contexts to share page-table mappings.
This makes it possible for user processes to run using different contexts while still sharing memory.
For CUDA 9.2, CUDA MPS is still invoked to keep shared allocations and communicate across them, but, with the exception of a handful of functions, most CUDA operations are performed without IPC communication.

Because sharing may serialize to use CUDA MPS, one slight disadvantage of CUDA IPC functions is that they have a measurable overhead. 
This can become a bottleneck. 
When sharing models at layer granularity, networks with large number of layers, such as {\textit ResNet269-v2}, have high overhead.
We remedy this by having a per-group of layer sharing or model sharing granularity.

The CUDA IPC overhead is measurable, and we can quantify whether using \nameofproject{} is beneficial statically using the empirical formula:  $\rho = b \div q - n\times(o + s)$, where $n$ is the number of objects to share (when the sharing granularity is at the model level, this value is $1$; when the granularity is at the layer, this value is the number of layers); $o$ is the overhead of sharing  CUDA memory via CUDA IPC and $s$ is the overhead of obtaining a CUDA device pointer from a shared CUDA IPC handle; $b$ is the number of bytes the model occupies on disk; and $q$ is the disk I/O bandwidth.
These constants can be computed once at system startup and cached to be used by \nameofproject{}.
If $\rho$ is positive, then its magnitude is correlated to the speedup one  gets using \nameofproject{}.
This equation can be used within the \nameofproject{} framework to determine at runtime whether to call \nameofproject{} to share a model or not and at what granularity to share the model.




%% file: sec/5-evaluation.tex

\section{Evaluation}\label{sec:evaluation}

\begin{table*}[t!]
    \centering
    \renewcommand{\arraystretch}{1.3}
    
    \resizebox{0.98\textwidth}{!}{%
\begin{tabular}{  | l | c | c | c | c | c | c | }
    \thead{Name} & \thead{CPU} & \thead{GPU} & \thead{Memory} & \thead{GPU Memory}  & \thead{Cached Reads}  & \thead{Buffered  Disk Reads}\\ \hline
    System 1 & Intel Core i9-7900X & TITAN Xp P110 & 32 GB & 12 GB & 8 GB/sec & 193.30 MB/sec \\
    System 2 & Intel Xeon E5-2698 v4& Tesla V100-PCIE & 256 GB & 16 GB & 10 GB/sec & 421.30 MB/sec \\
    System 3 & IBM S822LC Power8 w/ NVLink & Tesla P100-SXM2 & 512 GB & 16 GB & 27 GB/sec & 521.32 MB/sec \\
	\hline
\end{tabular}%
	}
	\caption{
    We evaluate \nameofproject{} on 3 systems which represent both cloud offerings and consumer desktop system configurations currently used for DL inference.
We use the Linux \texttt{hdparm} tool to measure the cached disk reads.
    }
	\label{tab:systems}
\end{table*}

We evaluate \nameofproject{} on 3 systems (shown in Table~\ref{tab:systems}) using $37$ (shown in Table~\ref{tab:models}) pre-trained small models and $8$ large models (shown in Table~\ref{tab:large_models}).
The systems selected represent  different types of instances that are currently provisioned in the cloud.
System 3 uses the NVLink  bus~\cite{foley2017ultra,tallent2017evaluating} which allows up to $35GB/s$ transfer between CPU and GPU.
System 3 is used as proxy for understanding our proposed method's behavior on high end cloud instances and next generation interconnects currently being deployed on HPC and cloud systems~\cite{vogt2017science,tharrington2017experiences}.
Multi-GPU results are similar to the single-GPU results shown bellow and for simplicity are omitted.

\begin{figure*}[htp]
  \centering
  \includegraphics[width=\textwidth]{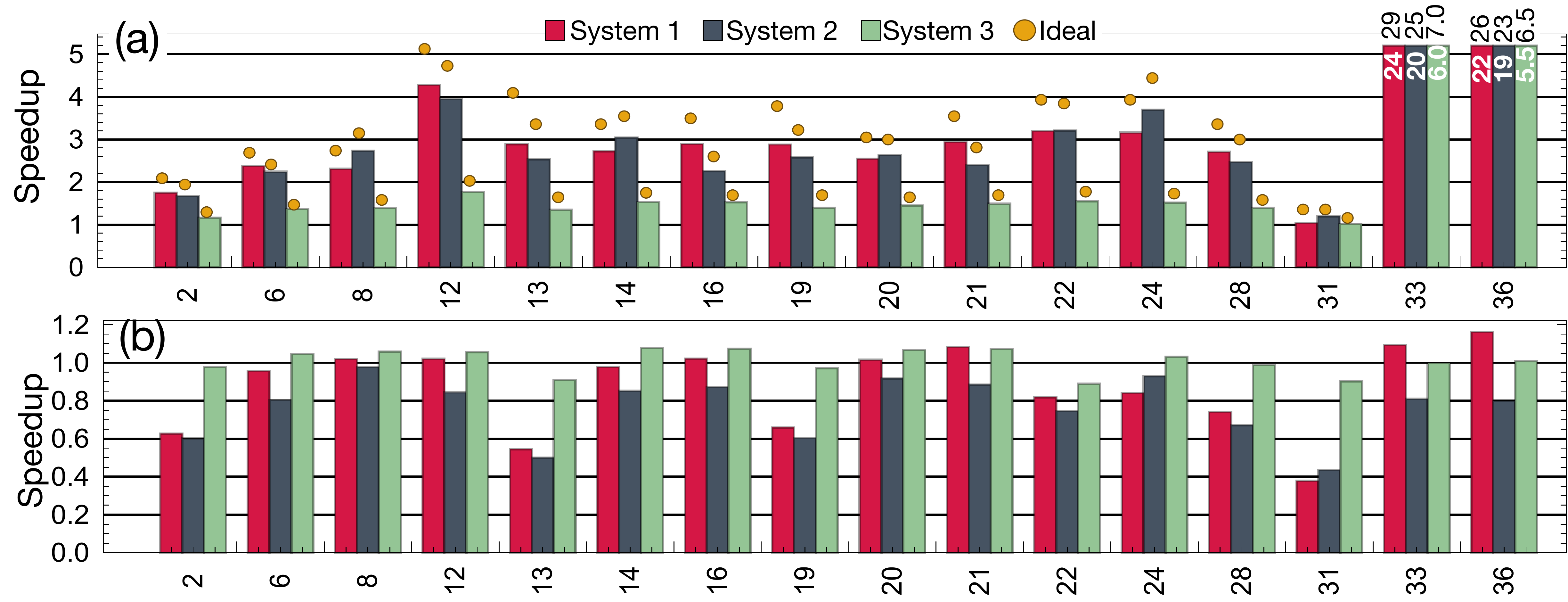}
  \caption{
  A representative sample of the models shown in Table~\ref{tab:models} are chosen and are run on the systems in Table~\ref{tab:systems} to achieve (a) the best case end-to-end time --- when the model has been pre-loaded in GPU memory --- and (b) the worst case end-to-end time --- when the model misses both the CPU and GPU persistence and needs to be loaded from disk.
  The speedups are normalized to end-to-end running time of the model without \nameofproject{}.
  The yellow dots show the ideal speedup; the speedup achieved by removing any I/O and data-transfer overhead --- keeping only the framework initialization and compute. 
  For models 33 and 36, the achieved speedup is shown on the bar (white) and the ideal speedup is shown on top of the bar (black).
  }
  \label{fig:speedup}
\end{figure*}

\begin{figure*}[tp]
  \centering
  \includegraphics[width=\textwidth]{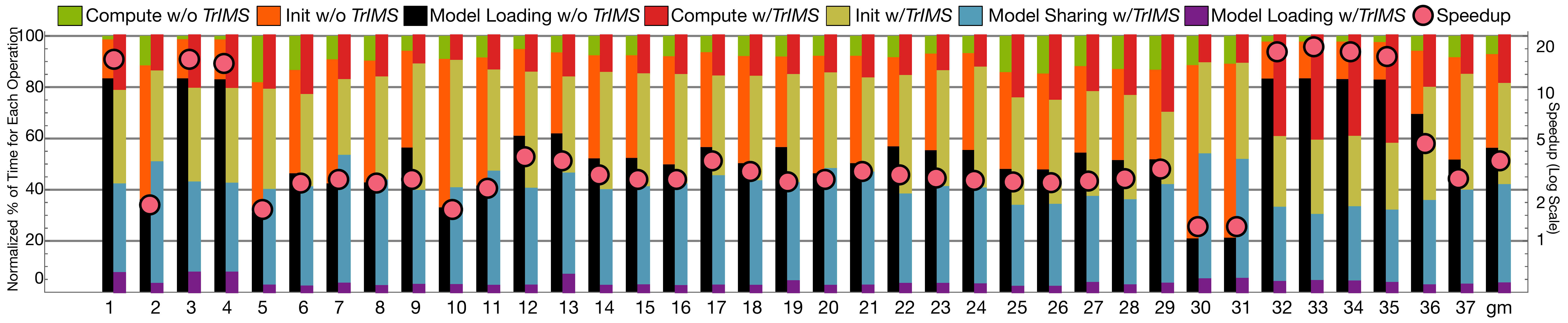}
  \caption{Detailed normalized times of operations with and without \nameofproject{} on System 3 using the models in Table~\ref{tab:models}. The duration for \nameofproject{} is normalized to the end-to-end time of not using \nameofproject{}.
  Model initialization is the time spent setting up the CUDA contexts for the model, initializing the the compute state, and (in the case of not using \nameofproject{}) copying the weights to GPU memory.
  Compute is the time spent performing inference computation.
  Model sharing is the overhead introduced by using \nameofproject{} and includes the gRPC communication and sharing GPU data using CUDA IPC.  
  Through \nameofproject{} we effectively eliminated model loading and data movement.
  }
  \label{fig:detailed_overhead}
\end{figure*}

\begin{figure}[tp]
  \centering
  \includegraphics[width=0.5\textwidth]{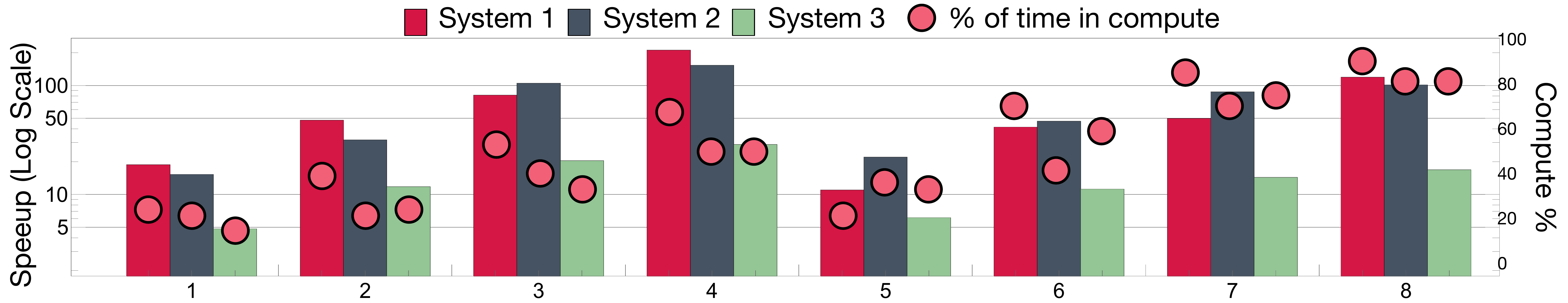}
  \caption{
   Large models in Table~\ref{tab:large_models} are run to achieve the best case end-to-end time --- when the model has been pre-loaded in GPU memory.
  The speedups are normalized to end-to-end running time of the model without \nameofproject{}.
  The red dots show the percentage of time spent performing the compute.
  We see linear speedup until the inference becomes compute bound.
  }
  \label{fig:large_speedup}
\end{figure}


We used image processing models as a representative workload because these are currently the most plentiful in FaaS pipelines.
\nameofproject{} is agnostic to the compute patterns of a network and the analysis would apply to other types of networks such as: RNNs, word embeddings, or matrix factorization.
The selected $37$ pre-trained image processing models, shown in Table~\ref{tab:models}, are based on their popularity in both research and usage. 
Some of the networks have variants.
These are used to simulate user trained models --- the same compute networks structure can have different weights.
Large models are used to show how \nameofproject{} performs with increasing  model sizes.

Throughout this section we compare our performance within a FaaS setting against ideal (where the model loading and data movement takes no time --- ideal is faster than model persistence) and use  end-to-end ``cold-start'' inference as the base line, since that's what is currently employed by FaaS environments.

\begin{table}[hb]
    \centering
    \resizebox{0.5\textwidth}{!}{%
\begin{tabular}{ | l | l | c | c | c | }
    \thead{ID} & \thead{Name} & \thead{\# Layers} & \thead{ILS}  & \thead{MWMF} \\ \hline
    1   &    AlexNet~\cite{krizhevsky2012imagenet}   &    16   &    516   &    238    \\
2   &    GoogLeNet~\cite{szegedy2015going}   &    116   &    111   &    27    \\
3   &    CaffeNet~\cite{krizhevsky2012imagenet}   &    16   &    512   &    233    \\
4   &    RCNN-ILSVRC13~\cite{girshick2014rich}   &    16   &    479   &    221    \\
5   &    DPN68~\cite{chen2017dual}   &    361   &    122   &    49    \\
6   &    DPN92~\cite{chen2017dual}   &    481   &    340   &    145    \\
7   &    Inception-v3~\cite{szegedy2016rethinking}   &    472   &    257   &    92    \\
8   &    Inception-v4~\cite{szegedy2017inception}   &    747   &    399   &    164    \\
9   &    InceptionBN-v2~\cite{ioffe2015batch}    &    416   &    313   &    129    \\
10   &    InceptionBN-v3~\cite{szegedy2016rethinking}   &    416   &    142   &    44    \\
11   &    Inception-ResNet-v2~\cite{szegedy2017inception}   &    1102   &    493   &    214    \\
12   &    LocationNet~\cite{weyand2016planet}   &    514   &    666   &    285    \\
13   &    NIN~\cite{lin2013network}   &    24   &    131   &    29    \\
14   &    ResNet101~\cite{he2016deep}   &    526   &    423   &    170    \\
15   &    ResNet101-v2~\cite{he2016deep}   &    522   &    428   &    171    \\
16   &    ResNet152~\cite{he2016deep}   &    777   &    548   &    231    \\
17   &    ResNet152-11k~\cite{he2016deep}   &    769   &    721   &    311    \\
18   &    ResNet152-v2~\cite{he2016deep}   &    761   &    340   &    231    \\
19   &    ResNet18-v2~\cite{he2016deep}   &    99   &    154   &    45    \\
20   &    ResNet200-v2~\cite{he2016deep}   &    1009   &    589   &    248    \\
21   &    ResNet269-v2~\cite{he2016deep}   &    1346   &    889   &    391    \\
22   &    ResNet34-v2~\cite{he2016deep}   &    179   &    222   &    84    \\
23   &    ResNet50~\cite{he2016deep}   &    268   &    270   &    98    \\
24   &    ResNet50-v2~\cite{he2016deep}   &    259   &    275   &    98    \\
25   &    ResNeXt101~\cite{xie2017aggregated}   &    526   &    375   &    170    \\
26   &    ResNeXt101-32x4d~\cite{xie2017aggregated}   &    522   &    378   &    170    \\
27   &    ResNeXt26-32x4d~\cite{xie2017aggregated}   &    147   &    147   &    59    \\
28   &    ResNeXt50~\cite{xie2017aggregated}   &    271   &    222   &    96    \\
29   &    ResNeXt50-32x4d~\cite{xie2017aggregated}   &    267   &    224   &    96    \\
30   &    SqueezeNet-v1.0~\cite{iandola2016squeezenet}   &    52   &    34   &    4.8    \\
31   &    SqueezeNet-v1.1~\cite{iandola2016squeezenet}   &    52   &    28   &    4.8    \\
32   &    VGG16~\cite{simonyan2014very}   &    32   &    1228   &    528    \\
33   &    VGG16-SOD~\cite{zhang2016unconstrained}   &    32   &    1198   &    514    \\
34   &    VGG16-SOS~\cite{zhang2015salient}   &    32   &    1195   &    513    \\
35   &    VGG19~\cite{simonyan2014very}   &    38   &    1270   &    549    \\
36   &    WRN50-v2~\cite{zagoruyko2016wide}   &    267   &    758   &    264    \\
37   &    Xception~\cite{chollet2016xception}   &    236   &    244   &    88 \\
    \hline
\end{tabular}%
	}
    \caption{The small models are popular models used in literature and is used as proxy models that offer a wide variety of sizes and computational complexity. Image classification models are used since they are the most commonly used.
    Both internal layer sizes (ILS) and the model weights memory footprint (MWMF) are shown in megabytes.
    The number of models is chosen to be 2x larger than the available 16 GB memory on Systems 2 and 3.
    }
  \setlength{\belowcaptionskip}{-10pt}
    \label{tab:models}
\end{table}

\subsection{Latency Improvement}\label{sec:speedup}

We measure the end-to-end ``cold-start'' inference of MXNet with and without \nameofproject{} -- for the sake of clarity we omit the input processing time.
Figure~\ref{fig:speedup} shows the achieved speedup on a representative set of the models compared against MXNet that does not utilize \nameofproject{}. We show two cases: (a) our best case (when there is a GPU cache hit) and (b) our worst case (when the cache misses both the CPU and GPU).

For best case analysis (Figure~\ref{fig:speedup}a), the server needs to create the CUDA IPC handles and the framework client needs to  embed the GPU device pointers within the framework's container. 
This introduces a slight overhead, however it is within $20\%$ of the ideal --- ideal defined as the time for inference where model loading or deserialization times set to zero.
We see that latency speedup improves proportionally to the model size, the system's data movement bandwidth, the system's compute resources, and the model's compute complexity.

For small models, where the I/O overhead is very low, for example SqueezeNet (which has a $5MB$ memory footprint), we observe only marginal speedup  ($1.04\times$).
These models are designed to have a small footprint --- targeting edge devices --- and are rarely used within the cloud. 
For state-of-the-art networks, such as VGG16-SOD, we observe $24\times$ speedup on System 1.
Even with fast disk and the NVLink interconnect, which mitigates I/O overhead by offering greater data movement bandwidth,  System 3 achieves $6\times$ speedup for VGG16-SOD. 

For the worst case analysis (Figure~\ref{fig:speedup}b), the MRM needs to load the data from disk, persist the model on the CPU, copy the data to the GPU, and send the GPU memory handles to the client.
Although we get a slow down, this case assumes there is no model sharing across pipelines, and therefore uncommon in  cloud setting.

\subsection{Speedup Breakdown}\label{sec:breakdown}


To understand where the new bottlenecks are for the inference using \nameofproject{}, we look at System 3 where we achieve the lowest speedup and measure the (a) time to perform inference computation, (b) time to initialize the model (this includes copying the data to the GPU when not using \nameofproject{}), (c) model loading from disk, and (d) model sharing overhead introduced by \nameofproject{}.
As can be seen in Figure~\ref{fig:detailed_overhead}, without using \nameofproject{} an average of $86\%$ of the time is spent loading and initializing the model while only $7\%$ is spent performing computation.
When using \nameofproject{} we eliminate the model loading from disk and remove the need to perform memory copies to the GPU.
Even though we introduce overhead, we still gain a $4.8\times$ geometric mean speedup.

\subsection{Large Model Evaluation}\label{sec:large_models}

We evaluate our method using large models which are common for medical image analysis, NLP, and time series modeling.
We generated the large models by starting with the regular AlexNet and VGG16 networks, keeping their compute graph topology, and rescaling the input dimensions to generate enlarged model.
Table~\ref{tab:large_models} shows the $8$ models selected for evaluation, their memory footprint, and their input sizes.

\begin{table}[htp!]
    \centering
    \resizebox{0.4\textwidth}{!}{%
\begin{tabular}{ | l | l | c | c | c | }
    \thead{ID} & \thead{Name} & \thead{Input Dims} & \thead{MWMF} \\ \hline
    1   &    AlexNet-S1~\cite{krizhevsky2012imagenet}   &   $227\times227$ &    238   \\
     2   &    AlexNet-S3~\cite{krizhevsky2012imagenet}   &   $454\times454$ &    770   \\
     3   &    AlexNet-S3~\cite{krizhevsky2012imagenet}   &   $681\times681$ &    1694   \\
     4   &    AlexNet-S4~\cite{krizhevsky2012imagenet}   &   $908\times908$ &  3010   \\
	 5   &    VGG16-S1~\cite{simonyan2014very}   &  $224\times224$ & 528  \\
	 6   &    VGG16-S2~\cite{simonyan2014very}   &  $448\times448$ & 1704 \\
	 7   &    VGG16-S3~\cite{simonyan2014very}   &  $672\times672$ & 3664 \\
	 8   &    VGG16-S4~\cite{simonyan2014very}   &  $896\times896$ & 6408 \\
    \hline
\end{tabular}%
	}
    \caption{Large models were used to evaluate our method. The models were generated by taking AlexNet and VGG16 and scaling the number of input features. 
    Large models arise in either medical image analysis, NLP, or time series analysis where down-sampling decreases the accuracy or the network requires a large window of features to give accurate results.
    }
    \label{tab:large_models}
  \setlength{\belowcaptionskip}{-10pt}
\end{table}

Figure~\ref{fig:large_speedup} shows that by removing model loading overhead, inference on large models becomes compute bound and gives an advantage to faster GPUs.
This is why System 1 achieves less speedup than System 2 for the more compute intensive VGG16 network (for example for model 7), since model inference computation accounts for $85\%$ on System 1 and $50\%$ on System 2. 
We expect this to be a more pronounced bottleneck for lower end GPUs and less of an issue for specialized low latency inference accelerators.

We also observe that \nameofproject{} increases the memory efficiency of the GPU.
Without \nameofproject{}, two inferences using model 8 cannot be run concurrently, since they overrun the GPU memory.
\nameofproject{} avoiding multiple private copies of the $6.4GB$ model on the same machine, enabling concurrent runs of large models.

\begin{figure*}[t!]
  \centering
  \includegraphics[width=0.95\textwidth]{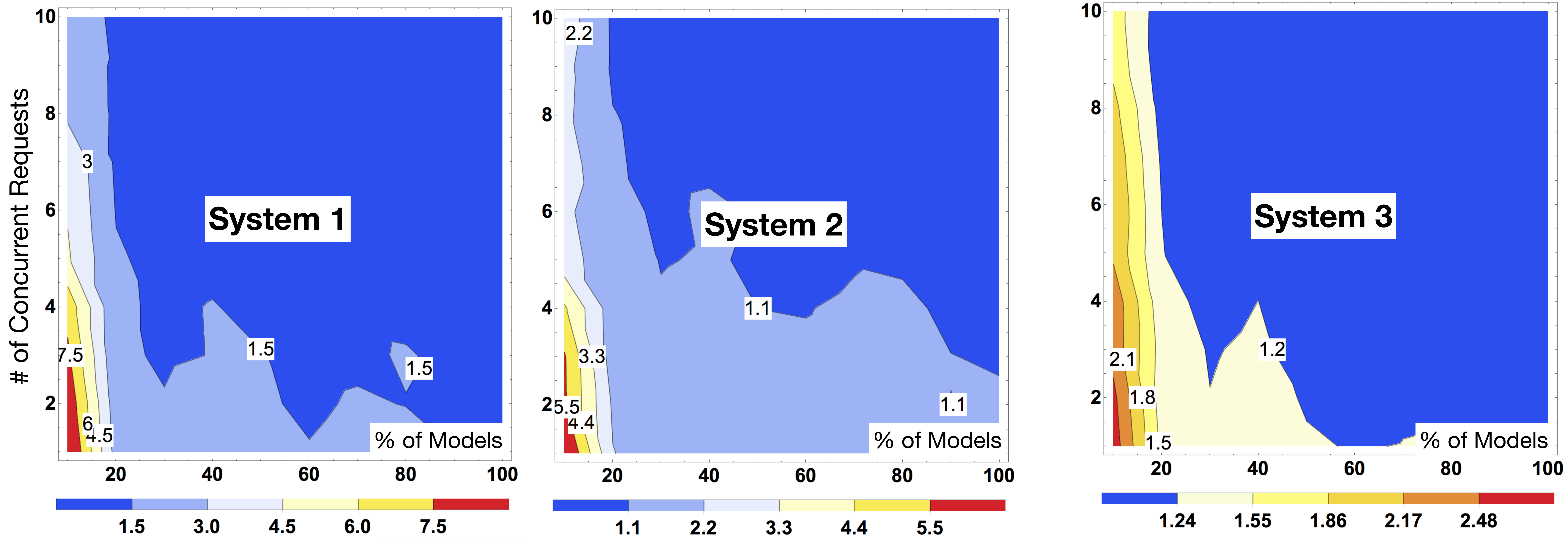}
  \caption{
 We vary the percentage of models run (from Table~\ref{tab:models}) and we sample them following a Pareto distribution (with $X=1$ and $l=1$). We also vary the concurrency level (number of inferences performed concurrently) ranging it from 1 to 10. The iso-curves show the geometric mean of the speedups for Systems 1, 2, and 3.
  }
  \label{fig:workload}
\end{figure*}

\subsection{Workload Modeling}\label{sec:workload}

Finally, we perform workload modeling to understand the behavior of \nameofproject{} on multi-tenant oversubscribed system.
The workload is selected from the 37 small models shown in Table~\ref{tab:models} following a Pareto distribution.
Since all the models cannot all be resident on the GPU at the same time --- in total having  $2\times$ GPU memory footprint --- the \nameofproject{} MRM needs to exercise the model reclamation and eviction procedure.
Because of limited space, we only present the results for the LRU eviction strategy, but our observations are valid for other eviction strategies.

Figure~\ref{fig:workload} shows the level iso-efficiency curves for the geometric mean speedup~\footnote{We measure the speedup value using the geometric mean across the $95\%$ latency speedup of each model.} as we vary the the concurrency level and number of models to run. 
We can see that even in an oversubscribed setting, we can still service $10$ clients concurrently,  reduce the overall batch execution time (by up to $8\times$), while incurring only a    $20\%$ latency penalty for each request. This slowdown is due to the cost of evicting models to accommodate the larger memory footprint, causing subsequent usage of the model miss the GPU cache.

For all three systems, we can observe an over-subscription sweet spot, where the percent of models and number of the concurrent request can be increased while the batch execution is preserved to a speedup of $1\times$. All systems show a sweet spot when $40\%$ of models are actively being requested. For system 1 and 3, the number of concurrent requests can be increased to 4, and system 2 the same number improves to 6. The difference in the over subscription sweet spot can be explained due to the different compute capabilities between the systems.
System 1 and 3 are provision with Pascal generation GPUs while system 2 has the latest Volta generation. Essentially, because we are successful in moving the inference bottleneck from I/O to compute, the  sweet spot is determined by the available computing resources.
In practice, cloud providers can perform  sensitivity analysis to determine the number of models hosted on each server and the number of concurrent requests to service based on the service's target requirements.

By removing model loading overhead, our speedups is bounded by the framework's inference pipeline.
Frameworks that are optimized for inference garner greater benefits.
For older generation or lower end GPUs, compute would likely dominate inference.
Therefore, if cloud providers are only interested in maintaining latency, they can utilize these older or lower end GPUs which have a lower initial cost of ownership.



%% file: sec/6-related.tex

\section{Related Works}\label{sec:related}

Recent related work has explored techniques to enable model serving at cloud scale. TesorFlow-Serving\cite{olston2017tensorflow} provides soft model isolation to guard against concurrent running request interfering with each other performance.
TFX\cite{baylor2017tfx} uses dedicated thread pool to hide model-loading overhead and provide thread-level user isolation.
Clipper\cite{crankshaw2017clipper} combine concurrent streams of DL requests into batches to better utilize the GPU at the cost of longer latency. 
All of these techniques suffer from their inability to provide user isolation or handle low latency inference.
 

Recent work~\cite{potluri2012optimizing,ji2012dma,pena2013evaluation,faraji2015hyper} leverage CUDA IPC in order to improve various intra-node and inter-node MPI collectives of a single process/application, and thus facilitate the porting to, and improve the performance of HPC applications on GPUs. MVAPICH2~\cite{mvapich2}, for instance, supports the use of MPI calls directly over GPU memory. Unlike these works, \nameofproject{} leverages CUDA IPC in order to persist data structures across processes and thus actively seeks to improve IO and memory footprint, instead of multi-GPU coordination.    

To reduce DL model inference latency and memory requirement, a large body of work have been performed recently in compacting and accelerating convolutional neural networks (CNNs). Quantization~\cite{gong2014compressing,wu2016quantized,vanhoucke2011improving,han2015deep,courbariaux2015binaryconnect} reduces the number of bits required to represent each weight by rescaling the weights to a domain smaller than the 32-bits required for floating point representation (usually 8-bits). Network pruning and sharing~\cite{srinivas2015data,han2015deep,chen2015compressing,zhou2016less} reduces redundant parameters which are not sensitive to the performance. Although these model optimization techniques can make the I/O vs. compute problem less severe, they have drawbacks and limited application scope. Quantized model inference suffer from accuracy loss while pruned network can significantly increase the computation intensity due to sparsity, especially on GPUs. Moreover these techniques currently only works with convolutional layers or fully connected layers, does not apply to other type of model inference, such as the fully connected layers used in word embedding. Model optimizations also could leverage \nameofproject{}, enabling the sharing of optimized CNN models.

Although various CPU/GPU virtualization techniques~\cite{morabito2015hypervisors,vgpu,duato2011enabling} and GPU multi-tenancy~\cite{cudamps,sengupta2013multi,yeh2017pagoda,ausavarungnirun2018mask} can improve system utilization and throughput through time sharing or parallel sharing CPU or GPU, they do not help solving the model loading overhead within inference processes. \nameofproject{} is orthogonal to these techniques and can be integrated into containers as a plugin. 
Also, in the very same way that NVIDIA Volta added the capability of effectively sharing memory across user-processes without the need of a proxy
process (CUDA MPS server).
The ability of sharing memory across different VMs (using a third level of virtual memory translation, as CPUs do)
would enable \nameofproject{} to work across VMs.



%% file: sec/7-conclusion.tex

\section{Conclusion}\label{sec:conclusion}

Collocating compute with model serving within FaaS overcomes the network barrier but suffers from high ``code start'' latency. 
We propose \nameofproject{} to mitigate the major source of ``code start'' latency --- the model loading overhead --- and make building complex latency sensitive pipelines with modular DL components feasible.
We do so by decoupling compute from model persistence and leveraging model sharing across user pipelines.
\nameofproject{} moves the bottleneck of DL model inference to compute, thus making GPU acceleration more appealing and making specialized novel inference architectures more tractable.

\nameofproject{}  was evaluated on three systems that represent current cloud system offerings. 
We used 45 DL models and show a speedup of up to $24\times$ for small models and up to $210\times$ for large models. When running concurrent inference, we can increase the overall system throughput by up to $8\times$. 
Our methodology, when applied to DL frameworks, offers advantages to both cloud providers and users. 
The isolation along with the significant memory reduction through model sharing enable cloud providers to over-provision hardware resources, thus decreasing the total cost of ownership. 
The benefits of \nameofproject{} to the cloud providers can be passed down to the users in the form of reducing latency or cost of inference. 

\nameofproject{} is a generic memory sharing technique that can be used when computation requires large number of constant parameters to be in situ on the CPU or GPU, while still maintaining isolation between users.
As such, the proposed method can be easily generalized to any application or algorithm that spans multiple processes and requires large amount of constant data resources.
While we motivated our work with deep learning, other types of applications such as image processing, physical simulation, or in-memory databases can benefit from our approach.




%% file: sec/9-ack.tex
\ignore{

\section{Acknowledgment}

This work is supported by IBM-ILLINOIS Center for Cognitive Computing Systems Research (C3SR) - a research collaboration as part of the IBM Cognitive Horizon Network.

}